\documentclass[twocolumn,preprintnumbers,superscriptaddress,amsmath,amssymb,aps,prl, 10pt]{revtex4-1}
\usepackage{physics}
\usepackage{blindtext}
\usepackage{graphicx}
\usepackage{xcolor}
\usepackage{amsmath}

\usepackage{amsthm}
\usepackage{amssymb}
\usepackage{amsfonts}
\usepackage{xr-hyper}
\usepackage{hyperref}
\hypersetup{
    colorlinks,
    linkcolor={blue},
    citecolor={blue},
    urlcolor={blue}
}
\usepackage{siunitx}
\usepackage{comment}
\usepackage{bm}
\usepackage{ulem,xpatch}
\usepackage{lineno}
\usepackage{indentfirst}
\usepackage{bbold}
\usepackage{todonotes}

\newcommand{\R}{\mathbb{R}}

\newcommand{\inn}{\mathrm{in}}
\newcommand{\outt}{\mathrm{out}}
\newcommand{\ext}{\mathrm{ext}}

\newcommand{\imu}{\text{\rm i}}
\newcommand{\expu}{\text{\rm e}}

\newcommand{\Gqba}{\Gamma_\text{q}}

\newcommand{\ain}{a_\text{in}}
\newcommand{\aout}{a_\text{out}}

\newcommand{\affil}{Photonics Laboratory, ETH Zürich, CH-8093 Zürich, Switzerland}
\newcommand{\affilQC}{Quantum Center, ETH Zurich, CH-8093 Zürich, Switzerland}
\newcommand{\affilICFO}{
ICFO -- Institut de Ciencies Fotoniques, The Barcelona Institute of Science and Technology, Castelldefels,
Barcelona 08860, Spain}
\newcommand{\affilICREA}{
ICREA -- Institucio Catalana de Recerca i Estudis Avancats, Barcelona 08010, Spain}
\newcommand{\affilTUWien}{
Institute for Theoretical Physics, Vienna University of Technology (TU Wien), 1040 Vienna, Austria}

\begin{document}

\title{Motional entanglement of remote optically levitated nanoparticles}

\author{N. Carlon Zambon}
\affiliation{\affil}
\affiliation{\affilQC}

\author{M. Rossi}
\altaffiliation{Present address: Kavli Institute of Nanoscience, Department of Quantum Nanoscience, Delft University of Technology, 2628CJ Delft, The Netherlands}
\affiliation{\affil}
\affiliation{\affilQC}

\author{M. Frimmer}
\affiliation{\affil}
\affiliation{\affilQC}

\author{L. Novotny}
\affiliation{\affil}
\affiliation{\affilQC}

\author{C. Gonzalez-Ballestero}
\affiliation{\affilTUWien}

\author{O. Romero-Isart}
\affiliation{\affilICFO}
\affiliation{\affilICREA}

\author{A. Militaru}
\altaffiliation{Present address: Institute of Science and Technology Austria, Am Campus 1, 3400 Klosterneuburg, Austria}
\affiliation{\affil}
\affiliation{\affilQC}

\begin{abstract}
  We show how to entangle the motion of optically levitated nanoparticles in distant optical tweezers. 
  The scheme consists in coupling the inelastically scattered light of each particle into transmission lines and directing it towards the other particle.
The interference between this light and the background field introduces an effective coupling between the two particles while simultaneously reducing the effect of recoil heating.
  We analyze the system dynamics, showing that both transient and conditional entanglement between remote particles can be achieved under realistic experimental conditions. 
\end{abstract}

\maketitle

Superposition states are one of the most fascinating manifestations of quantum mechanics. When dealing with two or more degrees of freedom, superpositions can produce strong correlations that make the joint state of the system non-separable, or entangled. Several works in the field of nanomechanics have prepared these quantum correlations between the motional degrees of freedom of two mechanical resonators, from individual atoms~\cite{Jost2009} 
to microbeams~\cite{Riedinger2018, Wollack2022}, microscale drum resonators~\cite{Ockeloen-Korppi2018, Kotler2021, deLepinay2021}, and acoustic modes of bulk resonators~\cite{Bienfait2020}. Extending this capability to levitated optomechanics~\cite{Millen2020, Gonzalez-Ballestero2021a}---i.e., generating motional entanglement between two optically levitated nanospheres in high vacuum---is a milestone in the field \cite{Rudolph2020,Chauhan2020,Brandao2021,Rudolph2023}. On the one hand, entangled states would allow levitated nanoparticles to show quantum motional features without necessarily requiring the preparation of non-Gaussian states, a task which remains challenging despite recent proposals~\cite{Martinetz2020, Neumeier2024, Roda-Llordes2024}. On the other hand, entangled states of two particles at controllable long distances could be used as probes to characterize yet unknown sources of decoherence~\cite{Miao2010}, as well as for quantum-enhanced sensing and metrology~\cite{Giovannetti2004, Giovannetti2004, Zhuang2018, Gessner2018, Gessner2018, Xia2023}.

Recent experiments have taken crucial steps towards entanglement in levitated optomechanics by showing mechanical ground-state cooling of levitated nanoparticles in free space~\cite{Tebbenjohanns2021, Magrini2021}, as well as strong and controllable light-mediated interactions between two levitated nanoparticles~\cite{Rieser2022, Reisenbauer2024}. In these setups, however, the trapping laser's shot noise induces a high degree of motional decoherence which prevents the generation of entanglement \cite{Rudolph2023}. So far, proposals to address this issue have included trapping nanoparticles inside a high-finesse optical cavity to enhance the coupling-to-decoherence ratio \cite{Rudolph2023, Delic2019, Windey2019, Vijayan2024}, non-optical coupling mechanisms~\cite{Rudolph2022, Winkler2024, Poddubny2024}, and using squeezed light to reduce measurement backaction noise \cite{Rudolph2023, Gonzalez-Ballestero2023}.

In this work we propose a method based on optical forces to generate entanglement between levitated nanoparticles across long distances (up to meter-scale) without a high-finesse optical cavity nor the use of squeezed light. We engineer long-range interactions by directional coupling of the light scattered off each nanoparticle into optical transmission lines within a closed loop configuration. Fine tuning of the accumulated phase in the transmission lines allows adjusting the effective coupling sign and strength, and to suppress the photon recoil. We derive the equations of motion for the system and provide analytical description of the system dynamics. Finally, we demonstrate the generation of both transient and conditional  entanglement.
\begin{figure}[t]
		\centering
		\includegraphics[trim=0cm 0cm 0cm 0cm, width=78mm]{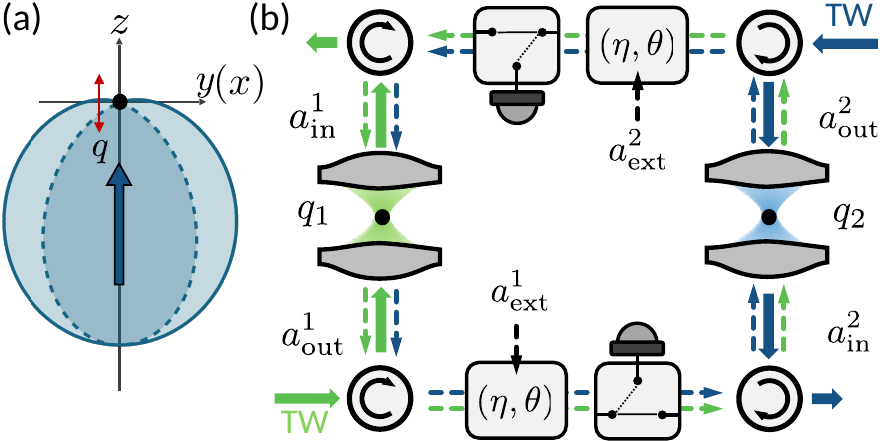}
		\caption{(a) Angular spectrum of the $q$ interacting mode associated with an $x$ polarized tweezer propagating along $z$. Solid (dashed) lines represent a cut in the $yz$ ($xz$) plane. (b) Schematic illustration of two optical tweezers interconnected via optical transmission lines characterized by a transmittance $\eta$ and phase delay $\theta$. The fields $\ain^j$ and $\aout^{j}$ denote the interacting modes prior and after interaction with each nanoparticle, $a^j_\text{ext}$  denotes the loss-induced vacuum fluctuations entering the loop.}
		\label{fig: setup}
\end{figure}

\textit{Model} - A dielectric nanoparticle illuminated by a tightly focused laser experiences a restoring optical force. For small displacements, its motion is harmonic and imprints a position-dependent phase onto the scattered laser photons, which in turn generate  recoil~\cite{Tebbenjohanns2019}. Since  scattering events occur randomly, recoil produces a fluctuating force acting on the nanoparticle. This form of optomechanical back-action represents the dominant fluctuating force in ultra-high-vacuum \cite{Jain2016}. 
Hereafter, we focus on the nanoparticle motion along the optical axis $z$ and introduce the displacement operator $q=z/(\sqrt{2}z_\text{zpf})$ normalized to the zero point fluctuations $z_\text{zpf}=\sqrt{\hbar/(2 m \Omega_0})$, where $m$ and $\Omega_0$ denote the nanoparticle mass and resonance frequency, respectively. 
While light-matter interactions in free space involve every mode of the electromagnetic continuum, it is possible to identify a collective mode that is solely responsible for the optomechanical interaction  with $q$: the interacting mode~\cite{Magrini2021,Militaru2022,Maurer2023}.
In the Heisenberg picture, the annihilation operators of the interacting mode before and after the interaction with the particle are denoted by $a_\inn$ and $a_\outt$, respectively. Crucially, the angular spectrum of the interacting mode associated with $q$ is strongly anisotropic: it predominantly propagates against the tweezer, see Fig.~\ref{fig: setup}(a).

Harnessing such directionality, we couple the back-scattered light, which carries information about the nanoparticle's motion, from the output of one tweezer to the input of another. This can be accomplished using circulators that define a one-way loop in combination with transmission lines (e.g., optical fibers). Hereafter, we consider the case of two identical optical tweezers interconnected with identical optical transmission lines, as illustrated in Fig.~\ref{fig: setup}(b). Each transmission line introduces a phase lag $\theta$, and has a finite transmittance $\eta$ owing to finite collection efficiency and imperfect mode matching. Losses, in turn, allow independent, uncorrelated modes $a_\ext$ to leak in the loop. Inline optical switches allow to sever the loop and divert the backscattered light from each nanoparticle to separate homodyne receivers that are used for state initialization and tomography.

We now derive the equations of motion for the composite system. As the transmission-line loop can be regarded as a bad ring cavity, delayed interactions within a mechanical period can be neglected if $\Omega_0 L/(\mathcal{F} c_g) \ll 1$ where $L$, $c_g$ and $\mathcal{F}$ denote the loop length, optical group velocity and cavity finesse, respectively. For standard optical fibers, using $\mathcal{F}=(1-\eta^2)/(1-\eta)^2$,  $\Omega_0 \approx 2\pi \times \SI{100}{kHz}$, and $\eta=0.5$, the approximation holds up to $L\sim 10~\mathrm{m}$~\cite{supplemental}. Each nanoparticle, labelled by the index $j=1,2$, obeys the Langevin equation 
\begin{equation}
    \label{eq: quantum Langevin}
    \ddot{q}_j + \Omega_0^2 q_j = \Omega_0 \sqrt{2\Gqba} \left( a_\inn^j + a_\inn^{j, \dagger} \right),
\end{equation}
where $\Gqba$ represents the decoherence rate due to quantum backaction. 
In addition to Eq.~\eqref{eq: quantum Langevin}, we can write the input-output relations~\cite{Gardiner1985}
\begin{equation}
\label{eq: input output}
a_\outt^j = a_\inn^j + \imu \sqrt{2\Gqba} q_j,
\end{equation}
with $\imu=\sqrt{-1}$ the imaginary unit. Equations~\eqref{eq: input output} and~\eqref{eq: quantum Langevin} show that the amplitude quadrature of the interacting mode drives the motion, while the phase quadratures probes it. Given the loop geometry considered in Fig.~\ref{fig: setup}(b), the input-output fields must satisfy the closure relations
\begin{equation}
\label{eq: loop branches}
a_\inn^j = \left( \sqrt{\eta} a_\outt^{3-j} + \sqrt{1-\eta}a_\ext^{3-j} \right) \expu^{\imu \theta}.
\end{equation}
\begin{figure}[t]
    \centering
	\includegraphics[trim=0cm 0cm 0cm 0cm, width=86mm]{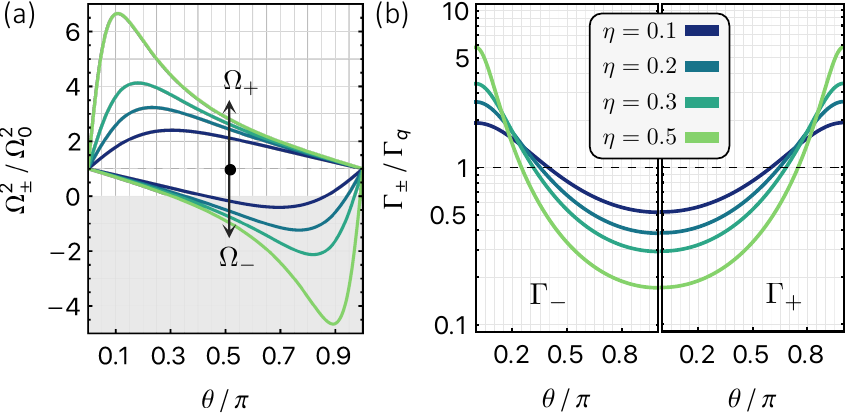}
    \caption{(a) Squared eigenfrequencies of the joint modes in units of the bare oscillator frequency $\Omega_0^2$ versus the transmission line phase ($\theta$), and for different collection efficiencies ($\eta$). In the shaded grey area the resonance frequency becomes imaginary, leading to an instability. (b) Dependence of the joint modes' decoherence rates $\Gamma_\pm$, in units of the bare oscillator one ($2\Gqba$). In both panels, the dashed black line traces the uncoupled oscillator case ($\eta=0$) and we used $\Gqba/\Omega_0=1$.}
    \label{fig: DOS and eigenfrequencies}
\end{figure}
Solving Eq.~\eqref{eq: input output} and~\eqref{eq: loop branches} for the two input fields yields~\cite{supplemental}
\begin{equation}\label{eq: loop solution}
    a_\inn^j = g_\mathrm{L} \left[ \imu \sqrt{2\Gqba} (\alpha q_j + q_{3-j}) + g_\eta (\alpha a_\ext^j + a_\ext^{3-j}) \right],
\end{equation}
where $\alpha(\theta)=\sqrt{\eta}\expu^{\imu \theta}$ denotes the transmission line transfer function, $g_\mathrm{L}(\theta) = \alpha/(1-\alpha^2)$ the Airy function of the effective low-finesse resonator generated by the loop, and $g_\mathrm{\eta}^2 = (1-\eta)/\eta$.
The amplitude quadrature of $a_\inn$ drives the nanoparticle's motion. Equation~\eqref{eq: loop solution} thus indicates that the loop effect is twofold. The first term, $\propto\alpha g_\mathrm{L}q_j$ renormalizes the trap stiffness depending on the round-trip phase $2\theta$. The second term, $\propto g_\mathrm{L}q_{3-j}$, introduces a coupling term originating from the modulation of the on-site optical force due to the interference between the tweezer and the interacting mode of the distant particle.  For convenience, we write dynamics in the normal mode basis upon introducing the joint modes $q_\pm = (q_1 \pm q_2)/\sqrt{2}$ and $p_\pm = (p_1 \pm p_2)/\sqrt{2}$, associated with the nanoparticle common $(+)$ and relative $(-)$ motion. From Eqs.~\eqref{eq: quantum Langevin} and~\eqref{eq: loop solution}, we obtain 
\begin{equation}
\label{eq: joint mode EOM}
\ddot{q}_\pm + \Omega_\pm^2 q_\pm = \Omega_0 \sqrt{4\Gqba} n_\pm,
\end{equation}
where
\begin{equation}\label{eq: normal mode splitting}
    \Omega_\pm^2=\Omega_0\left(\Omega_0 \pm\frac{4\Gqba\sqrt{\eta}\sin(\theta)}{1+\eta\mp2\sqrt{\eta}\cos(\theta)}\right)   
\end{equation}
defines the normal mode frequencies, and $n_\pm$ are two mutually uncorrelated and non-local fluctuations driving the joint modes. The derivation of Eq.~\eqref{eq: joint mode EOM} and the expressions for $n_\pm$ are provided in~\cite{supplemental}. 

In Fig.~\ref{fig: DOS and eigenfrequencies}(a) we plot $\Omega_\pm^2$ for $\Gqba/\Omega_0=1$ as a function of the phase $\theta$. For $\theta \in \{0, \pi \}$ we observe that the value of $\Omega_+^2$ ($\Omega_-^2$) is always larger (smaller) than the bare oscillator one $\Omega_0^2$. This fact is a manifestation of the normal mode splitting in coupled oscillators. Note that the splitting can exceed the frequency bare modes, which is the hallmark of ultra-strong coupling \cite{Forn-Diaz2019,FriskKockum2019a,Markovic2018}. Consequently, $\Omega_-^2$ becomes negative and the anti-symmetric mode $q_{-}$ becomes unstable (shaded gray area)~\cite{supplemental}. Finally, we calculate the decoherence rates $\Gamma_{\pm}=2\Gqba\mathcal{N}_\pm^2$,which are proportional to the effective photon recoil rates that drive the joint modes. We obtain
\begin{equation}
       \mathcal{N}_\pm^2 =\int_\R\dd t\,\langle n_\pm(t) n_\pm(t')\rangle=\frac{1}{2}\frac{1-\eta}{1+\eta\mp2\sqrt{\eta}\cos(\theta)}.
\end{equation}
Figure~\ref{fig: DOS and eigenfrequencies}(b) shows the dependence of $\Gamma_{\pm}$ on the  phase $\theta$ and transmittance $\eta$. While the noise strength for the common mode ($+$) is peaked at $\theta=0$ and reaches a minimum at $\theta=\pi$, the trend is opposite for the relative mode ($-$). Importantly, it is possible to identify values of $\theta$ where both $\Gamma_{+}$ and $\Gamma_{-}$ are lower than the corresponding bare oscillator value (dashed black line). Thus, the loop effect is not just to redistribute photon recoil between the two particles but can actually reduce it overall. We can understand this fact by noticing that the loop introduces an effective low-finesse resonator (limited by $\eta$), which in turn suppresses the density of states into which the particle scatters. All panels in Fig.~\ref{fig: DOS and eigenfrequencies} can be extended to the phase interval $[\pi, 2\pi]$ by swapping the common and relative modes.\\ 

\textit{Dynamics}- the expectation value and covariance matrix of the state vector $\vb{v}^{T} = (q_+, p_+, q_-, p_-)$ fully encode the state. Initially, we consider both nanoparticles to be in a low-occupation state with $\langle\vb{v}_0\rangle=0$ and covariance matrix $\bm{\Sigma}_0 = \langle \overline{\vb{v}_0 \vb{v}_0^\mathrm{T}} \rangle$, where the overline denotes symmetrization. State preparation can be accomplished by using the optical switches to reroute the scattered light from each particle to separate homodyne receivers, see Fig~\ref{fig: setup}(b). The measurement records are then used to stabilize the nanoparticle conditional state using feedback \cite{Mancini1998,Meng2020,Isaksen2023}.

Starting from this initial condition, we compute the evolution of the covariance matrix $\vb{\Sigma}=\vb{\Sigma}^{c}+\vb{\Sigma}^{n}$, which is split into two terms associated with the coherent and incoherent dynamics, respectively. Since the joint modes in Eq.~\eqref{eq: joint mode EOM} are decoupled, each diagonal block of $\bm{\Sigma}$ evolves independently. The matrix exponential generating the flow of each subspace is $\vb{\Phi}_{\pm}(t)=\vb{S}[\sqrt{r_{\pm}}] \vb{R}[\phi_\pm] \vb{S}[\sqrt{r_{\pm}}]^{-1}$, where $\vb{S}[\cdot]$ is a squeezing matrix with parameter $\sqrt{r_{\pm}}=\sqrt{\Omega_0/\Omega_\pm}$, and $\vb{R}[\phi_\pm]$ is a clockwise rotation by an angle $\phi_{\pm}=\Omega_{\pm} t$ in phase-space. The coherent terms yield
\begin{equation}\label{eq: coherent evolution sigma}
	\vb{\Sigma}^c_{\pm}(t)=\vb{\Phi}_{\pm}(t)\vb{\Sigma}_0\vb{\Phi}_{\pm}(t)^T.
\end{equation}
For stable dynamics, i.e., $\Omega_-^2>0$, each element in Eqs.~\eqref{eq: coherent evolution sigma} oscillate at twice $\Omega_\pm$. In contrast, in the case of unstable dynamics, the covariance matrix elements get squeezed at a rate $\propto\exp(2\Omega_\pm t)$ \cite{Romero-Isart2017}. The incoherent contribution is
\begin{equation}
\label{eq: incoherent evolution sigma}
	\vb{\Sigma}^n_{\pm}(t)=\frac{r_{\pm}^2\Gamma_{\pm}}{\Omega_{\pm}}\begin{pmatrix}
		\phi_{\pm}-\frac{\sin2\phi_{\pm}}{2} & r_{\pm}^{-1}\sin^2\phi_{\pm}\\
		r_{\pm}^{-1}\sin^2\phi_{\pm} & \frac{2\phi_{\pm}+\sin2\phi_{\pm}}{2r_{\pm}^2}\\
	\end{pmatrix},
\end{equation}
\begin{figure}[t]
    \centering    
    \includegraphics[trim=0cm 0cm 0cm 0cm, width=86mm]{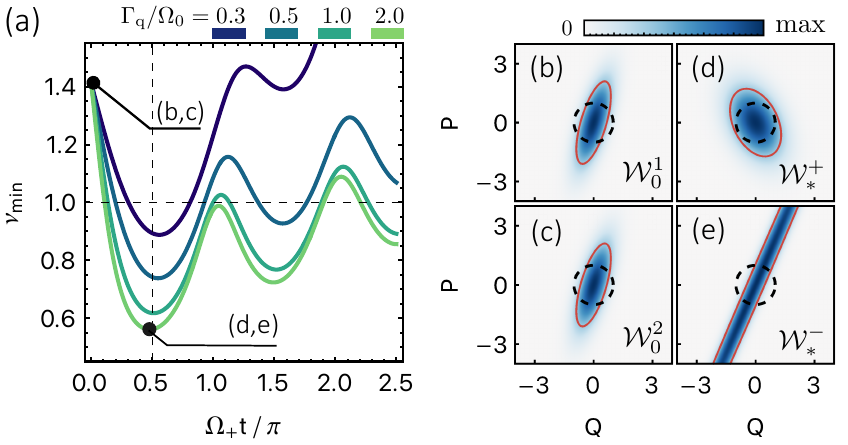}
    \caption{(a) Time evolution of the entanglement witness $\nu_\text{min}$ for different values of $\Gqba$. The loop efficiency is $\eta = 0.5$ and the phase  $\theta = 2\pi/3$. The entanglement revivals oscillate at twice $\Omega_+$, since $q_-$ is unstable. (b,c) Wigner distributions of the two particles at $t=0$, corresponding to the initial state of the bare oscillators for $\Gqba/\Omega_0=2$. The dashed circle is the covariance of the zero point motion. (d,e) Wigner distributions of the joint modes after an interaction time $t^*\approx \pi/(2\Omega_+)$, showing characteristic two-mode squeezing features.
    }
    \label{fig: transient log negativity}
\end{figure}
indicating a monotonous growth of the position and momentum variance at a rate $\Gamma_\pm$, and correlation oscillations at $2\Omega_\pm$~\cite{supplemental}. In the following, we use our knowledge of the time-dependent covariance matrix $\vb{\Sigma}(t)$ to demonstrate that the in-loop dynamics generates motional entanglement.

\textit{Transient entanglement} - According to the Duan-Simon criterion, the motional state of the two particles is entangled if $\nu_U = \langle \Delta q^2_{+,U} \rangle + \langle \Delta p^2_{-,U}\rangle < 1$ for some local symplectic transformation $U$ \cite{Simon2000,Duan2000}. We can generalize this statement by introducing $\nu_\mathrm{min}$, the minimum value of $\nu_U$ under all possible symplectic transformations $U$ \cite{Serafini2004,Weedbrook2012,Kotler2021}. Figure~\ref{fig: transient log negativity}(a) shows the time evolution of $\nu_\mathrm{min}(t)$ for a loop efficiency $\eta=0.5$, a phase $\theta = 2\pi/3$, and for some representative values of  $\Gqba/\Omega_0$. In all curves $\Omega_{-}$ is imaginary, while the common mode (+) is stable. As  a result, $\nu_\mathrm{min}$ oscillates at twice the stable mode frequency $\Omega_{+}$. The separability criterion is thus maximally violated at a time $t^* \sim \pi/(2\Omega_{+})$. Moreover, $\nu_\text{min}$ decreases with increasing $\Gqba$. This is because the correlations (two-mode squeezing) scale exponentially with $\Omega_{-}$, which in turn grows with $\Gqba$ according to Eq.~\eqref{eq: normal mode splitting}. In contrast, the decoherence rate scales only linearly with $\Gqba$. Finally, the entanglement vanishes at large times as photon recoil eventually degrades the initial state purity.

Figure.~\ref{fig: transient log negativity}(b-e) shows the Wigner functions $\mathcal{W}_{j} \sim \text{exp}[-\vb{v}_{j}^\mathrm{T} \vb{\Sigma} \vb{v}_{j}]$ of the initial state in the single particle basis $\mathcal{W}_{0}^{1,2}$, and after an interaction time $t^*$ in the joint mode basis $\mathcal{W}_{*}^{\pm}$, for $\Gqba/\Omega_0=2$. Position squeezing in Fig.~\ref{fig: transient log negativity}(b,c) is due to the departure from the weak measurement limit ($\eta\Gqba/\Omega_0\ll 1$) and the consequent break down of the rotating-wave approximation \cite{Meng2020}. Moreover, we notice that the covariance ellipses (solid tangerine lines) associated with the Wigner functions $\mathcal{W}_{*}^{\pm}$ are anti-correlated and that the unstable mode (-) is squeezed $7.5~\mathrm{dB}$ below the zero point motion (dashed black lines) at an angle $\xi_{-}=\arg(1-r_{\pm}^{-1})$. These are both signatures of the emergence of large two-mode squeezing interactions in ultra-strongly coupled oscillators \cite{Markovic2018,Kustura2022}.
\begin{figure}[t]
     \centering    
    \includegraphics[trim=0cm 0cm 0cm 0cm, width=80mm]{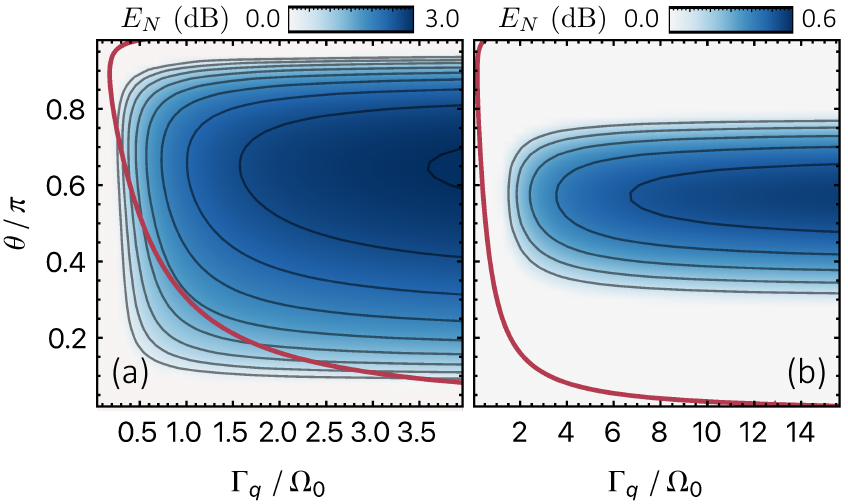}
    \caption{Maximal log negativity generated by  transient dynamics versus $\Gqba$ and $\theta$ . Panel (a) and (b) correspond to a loop efficiency $\eta = 0.5$ and $\eta = 0.3$, respectively. The solid red line bounds from the right the instability region.}
    \label{fig: optimum log negativity}
\end{figure}

We conclude this section extracting the maximal logarithmic negativity $E_N=-10\text{min}[0,\log_{10}(\nu_\text{min})]$ as a function of $\theta$ and $\Gqba$. We show the results in Fig.~\ref{fig: optimum log negativity} for the case of $\eta=0.5$ [panel (a)] and $\eta=0.3$ [panel (b)]. Interestingly, even if larger negativities are reached within the unstable region in parameter space (solid red line), for $\eta=0.5$ the particles are entangled  even in the stable portion of parameter space. Moreover, the negativity for a fixed $(\eta,\theta)$ first grows with $\Gqba$ but finally saturates for $\Gqba/\Omega_0\gg 1$. Indeed, in such a regime $\Omega_{\pm}^2\propto \pm \Gqba$ but as correlations grow $\propto \exp(\Omega_{-} t)$, the optimal interaction time $t^*\propto\Omega_{+}^{-1}$ decreases, resulting in a squeezing factor $\exp(2\Omega_{-}\Omega_{+}^{-1})$ independent of $\Gqba$. 

To experimentally certify the generation of an entangled state of motion, one needs to reconstruct $\vb{\Sigma}$. As we show in~\cite{supplemental}, this can be accomplished via an optimal retrodiction filter after subtracting the imprecision noise associated with the monitoring process \cite{Wiseman2009,Wieczorek2015,Meng2020}.


\textit{Stationary conditional state entanglement} - In the previous sections we have considered a binary situation where all the back-scattered light was either circulating in the loop or was diverted into homodyne receivers. We now investigate an intermediate configuration where a fraction $\eta_\text{m}$ of the light coupled in the transmission line (with collection efficiency $\eta_c$) is used to measure the system. Simultaneously, a fraction $\eta=\eta_c(1-\eta_\mathrm{m})$ circulates in the loop. Our goal is to show that the conditional state of the system is also non-separable.

\begin{figure}[t]
		\centering
		\includegraphics[trim=0cm 0cm 0cm 0cm, width=86mm]{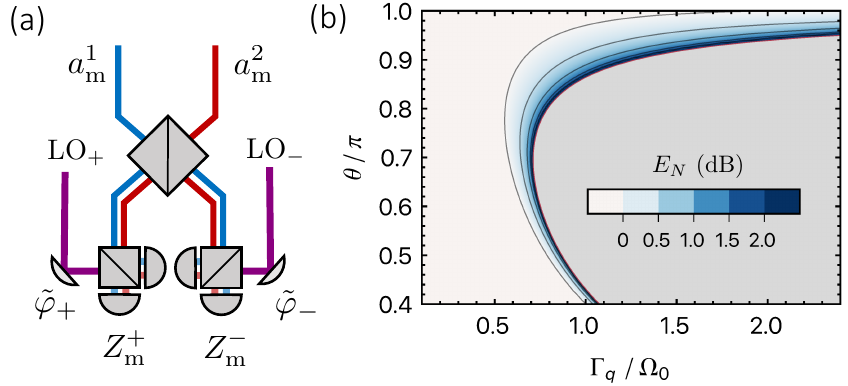}
		\caption{(a) Schematic illustration of the joint modes displacement measurement apparatus. (b) Log negativity of the conditional state as a function of $\Gqba$ and $\theta$. The collection and measurement efficiency are $\eta_c = 0.5$ and $\eta_\mathrm{m} = 0.8$, respectively. The gray region indicates unstable solutions.
  }\label{fig:Conditional_Entanglement}
\end{figure}
We denote with $a_\text{m}^{1,2}$ the optical fields at the measurement ports. In-loop correlations among the noise terms $a_\mathrm{ext}^i$ makes the analytic expression of $a_\mathrm{m}^{1,2}$ rather lengthy; we report it in~\cite{supplemental}. The key feature is that combining the two fields with a beam splitter, as shown in Fig.~\ref{fig:Conditional_Entanglement}(a), results in output fields $a_\mathrm{m}^\pm = (a_\mathrm{m}^1 \pm a_\mathrm{m}^2)/\sqrt{2}$ that encode only information about the respective joint mode displacement $q_\pm$. A homodyne receiver with analyzer angle $\varphi$ measures the quadrature $Z_\text{m}^{\pm}=\cos\varphi X_\text{m}^{\pm}+\sin\varphi Y_\text{m}^{\pm}$, where $X_\text{m}^{\pm}$ and $Y_\text{m}^{\pm}$ denote respectively the amplitude and phase quadratures of the fields $a_\mathrm{m}^\pm$. Interestingly, we showed that fixing the analyzer angle $\tilde{\varphi}_{\pm}$ of each detector to 
\begin{equation}
	\tan\tilde{\varphi}_\pm=\cot(\theta)\mp [\sqrt{\eta}\sin(\theta)]^{-1}
\end{equation}
allows collecting the maximal amount of information on $q_{\pm}$, with efficiency $\tilde{\eta}=\eta_c\eta_\mathrm{m}/(1-\eta)$. Correspondingly, the imprecision noise inherent to the measurement and the process noise (back-action) decorrelate, allowing to map our system onto the standard problem of continuous position measurements in optomechanics \cite{Wiseman2009,doherty_quantum_2012}. Drawing from its toolbox, we can readily write the conditional state covariance matrix $\vb{\Sigma}_{W}$ of the bipartite system~\cite{supplemental}. 

In Fig.~\ref{fig:Conditional_Entanglement}(b) we calculate $\nu_\text{min}$ for the conditional state covariance $\vb{\Sigma}_{W}$ as a function of the back-action rate $\Gqba$ and transmission line phase $\theta$. We assume a collection efficiency $\eta_c=0.5$, and measured fraction $\eta_\mathrm{m}=0.8$, yielding effectively a loop transmission $\eta=\eta_c(1-\eta_\mathrm{m})=0.1$. In some regions of the parameter space, especially in vicinity of the unstable region, $\nu_\text{min}$ reaches values that are comparable to those obtained in the transient dynamics, even for moderate ratios $\Gqba/\Omega_0\approx 1$. We therefore anticipate that the measurement outcomes at the optimal analyzer angles $\tilde{\varphi}_\pm$ can be processed using an optimal filter and controller to asymptotically stabilize the conditional state, thereby preparing an entangled steady state of motion of the two nanoparticles.\\

\textit{Conclusion}- We proposed a scheme  
to entangle levitated nanoparticles held in optical tweezers at meter-scale distances solely harnessing optical forces. Importantly, our scheme does not rely on high-finesse cavities or the injection of squeezed light.
The nanoparticles must be trapped in ultra-high-vacuum where their motion is predominantly driven by photon recoil, an already demonstrated regime~\cite{Jain2016}. Coupling the backscattered light into the loop is equivalent to maximizing the detection efficiency~\cite{Tebbenjohanns2019}, where values higher than 30\% have been achieved~\cite{Magrini2021}. Finally, the phase acquired in the loop can be stabilized by extracting and monitoring a small fraction of the circulating light.

Generalizing our results to an asymmetric configuration, e.g. by setting uneven transmission line phases, will feature a rich parameter space characterized by non-reciprocal interactions and vacuum noise correlations \cite{Rudolph2023}.
Such entangled states may enhance the force-gradient sensing capabilities of our platform \cite{Rudolph2022}, with  applications in searches of new physics \cite{afek_coherent_2022}. Moreover, entangling the motion of massive objects at large distances is a promising prospect for testing quantum mechanics~\cite{Ghirardi1986,Bassi2013,Arndt2014,Gonzalez-Ballestero2021a} or to perform locality loophole-free Bell tests with levitated objects \cite{Ralph2000,Thearle2018}, a task that is significantly more challenging as it requires non-Gaussian operations~\cite{Banaszek1998, Stobinska2007}.

\begin{acknowledgments}
\section{Acknowledgements}
This research has been supported by the European Research Council (ERC) under the grant agreement No. [951234] (Q-Xtreme ERC-2020-SyG), the Swiss SERI Quantum Initiative (grant no. UeM019-2), and the Swiss National Science Foundation (grant no. 51NF40-160591). N.C-Z. thanks for support through an ETH Fellowship (grant no. 222-1 FEL-30). The research was funded in part by the Austrian Science Fund (FWF) [10.55776/COE1]. For Open Access purposes, the author has applied a CC BY public copyright license to any author accepted manuscript version arising from this submission.
\end{acknowledgments}

\bibliographystyle{apsrev4-1}

%

\end{document}


\title{Supplemental Material: \\ Motional entanglement of remote optically levitated nanoparticles}

\author{N. Carlon Zambon}
\affiliation{\affil}
\affiliation{\affilQC}

\author{M. Rossi}
\altaffiliation{Present address: Kavli Institute of Nanoscience, Department of Quantum Nanoscience, Delft University of Technology, 2628CJ Delft, The Netherlands}
\affiliation{\affil}
\affiliation{\affilQC}

\author{M. Frimmer}
\affiliation{\affil}
\affiliation{\affilQC}

\author{L. Novotny}
\affiliation{\affil}
\affiliation{\affilQC}

\author{C. Gonzalez-Ballestero}
\affiliation{\affilTUWien}

\author{O. Romero-Isart}
\affiliation{\affilICFO}
\affiliation{\affilICREA}

\author{A. Militaru}
\altaffiliation{Present address: Institute of Science and Technology Austria, Am Campus 1, 3400 Klosterneuburg, Austria}
\affiliation{\affil}
\affiliation{\affilQC}

\maketitle

\tableofcontents
\clearpage

\section{Input fields }\label{app: input fields}

In order to obtain Eq.~(4) from the main text, one can start by imposing the in-loop conditions
\begin{equation}
    \begin{aligned}
        \ain^{1} & =\expu^{\imu \theta_2}(\sqrt{\eta_2}\aout^{2}+\sqrt{1-\eta_2}a_\text{ext}^2),\\
         \ain^{2} & =\expu^{\imu \theta_1}(\sqrt{\eta_1}\aout^{1}+\sqrt{1-\eta_1}a_\text{ext}^1).
    \end{aligned}
\end{equation}
Inserting the input-output relations [Eq.~(2)] yields
\begin{equation}
    \begin{aligned}
        \ain^{1} & =\expu^{\imu \theta_2}(\sqrt{\eta_2}(\ain^{2}+\imu \sqrt{2\Gqba}q_2)+\sqrt{1-\eta_2}a_\text{ext}^2),\\
         \ain^{2} & =\expu^{\imu \theta_1}(\sqrt{\eta_1}(\ain^{1}+\imu \sqrt{2\Gqba}q_1)+\sqrt{1-\eta_1}a_\text{ext}^1).
    \end{aligned}
\end{equation}
Finally one can solve this equation set for the input amplitudes. Defining $\alpha_j=\sqrt{\eta_j}\expu^{\imu \theta_j}$ and inserting the second relation in the first we get
\begin{equation}
    \begin{aligned}
    &(1-\alpha_1\alpha_2)\ain^{1}=\imu \sqrt{2\eta_2\Gqba} \expu^{\imu (\theta_1+\theta_2)}(\sqrt{\eta_1}q_1+\expu^{-\imu \theta_1}q_2) 
    + \expu^{\imu(\theta_1+\theta_2)}(\sqrt{\eta_1(1-\eta_2)}a_\text{ext}^{1}+\expu^{-\imu\theta_1}\sqrt{1-\eta_2}a_\text{ext}^{2}),
    \end{aligned}
\end{equation}
which can be simplified into Eq.~(4) assuming a fully symmetric configuration, i.e. $\eta_{1}=\eta_{2}=\eta$ and $\theta_{1}=\theta_{2}=\theta$, and introducing the short hand notations $g_\mathrm{L} = \alpha/(1-\alpha^2)$ and $g_\mathrm{\eta}^2 = (1-\eta)/\eta$.\\

Next we analyze the optical density of states (DOS) in the coherent loop. To do this, it is sufficient to evaluate the input field variance associated to the leaky external fields $a_\text{ext}^{j}$. We do so in the simplest fully symmetric case. The noise terms associated with information loss in the loop read $a_{n}=g_L g_{\eta} (\alpha a_\text{ext}^{j}+\alpha a_\text{ext}^{j-3})$, and their correlation is then
\begin{equation}\label{eq: opticalDOS}
    \begin{aligned}
        \langle a^{j}_{n}(t)a^{j}_{n}(t')^{\dag}\rangle &=\frac{1-\eta^2}{1-2\eta\cos 2\theta + \eta^2} \,\delta(t-t') = \rho^{\eta,\theta}_{\text{opt}} \delta(t-t').
    \end{aligned}
\end{equation}
All other two point correlators vanish under the assumption that the external fields leaking into the loop are in vacuum. Interestingly, the above equation corresponds to the Airy function modeling the transmission of a resonator, with $\theta$ representing the round-trip phase spanning $[0,\pi]$ within a free spectral range. Eq.~\eqref{eq: opticalDOS} describes how the presence of the loop modifies the DOS. Depending on the phase factor $\theta$, the DOS can be either boosted or suppressed, which reflects directly into the strength of the recoil heating driving each particle.\\

We now calculate the input field amplitude quadrature, which is the one driving the motion of the particles. Hereafter we assume that the system is fully symmetric, and we focus on $X_\text{in}^{1} = (\ain^1 + \ain^{1,\dag})/\sqrt{2}$ as the expression for $X_\text{in}^{2}$ follows imposing exchange symmetry. We can separate the amplitude quadrature terms in a coherent contribution
\begin{equation}\label{eq: SPB_Xin_coherent}
\begin{aligned}
    X_\text{in}^{1,c} & = -\sqrt{2\Gqba}\text{Im}[ g_L( \alpha q_1 +  q_2)]\\
        &=-\sqrt{4\Gqba} \left[ f_1(\theta) q_2 + g_{1}(\theta) q_1 \right],
\end{aligned}
\end{equation}
plus a noise term associated with vacuum fluctuations
\begin{equation}\label{eq: SPB_Xin_noise}
\begin{aligned}
	 X_\text{in}^{1,n} &=\text{Re}[g_{\eta}g_L (\alpha a_\text{ext}^1+a_\text{ext}^2)]\\
	 & =g_\eta[(g_2 X_\text{ext}^1+f_2 X_\text{ext}^2)-(g_1 Y_\text{ext}^1 + f_1 Y_\text{ext}^2)],
\end{aligned}    
\end{equation}
where we have introduced the coefficients 
\begin{equation}\label{eq: Xin_incoherent}
    \begin{aligned}
        f_1(\theta) & = \frac{\sqrt{\eta}(1+\eta)\sin\theta}{1-2\eta\cos 2\theta+\eta^2},\\
        f_2(\theta) & = \frac{\sqrt{\eta}(1-\eta)\cos\theta}{1-2\eta\cos 2\theta+\eta^2},\\
        g_1(\theta) & = \frac{\eta \sin2\theta}{1-2\eta\cos 2\theta+\eta^2},\\
        g_2(\theta) & = \frac{\eta(\cos2\theta-\eta)}{1-2\eta\cos 2\theta+\eta^2}.\\
    \end{aligned}
\end{equation}
By inspection, Eq.~\eqref{eq: SPB_Xin_coherent} amounts to a bilinear coupling between the two particles (first term), and a self-energy contribution that modifies the resonance frequency of each particle (second term). The fluctuating force in Eq.~\eqref{eq: SPB_Xin_noise} drives the motion of the coupled system and depends upon the collection efficiency $\eta$ and phase $\theta$ of the transmission lines, as one would expect given the modified DOS in the loop, cf. Eq.~\eqref{eq: opticalDOS}. Furthermore, these optical fluctuating forces are mutually correlated. Their variance and correlation are
\begin{subequations}
    \begin{align}
        \langle (X^{j,n}_\text{in})^2 \rangle &= \frac{1}{2}\rho_\text{opt}^{\eta,\theta},\\
        \langle \overline{X^{1,n}_\text{in}X^{2,n}_\text{in}}\rangle &= \frac{\sqrt{\eta}\cos(\theta)}{1-\eta^2}\rho_\text{opt}^{\eta,\theta},
    \end{align}
\end{subequations}
which we computed assuming once again an input vacuum state. Both quantities are proportional to the density of optical states in the loop, while correlations among the baths driving the particle motion vanish only for specific values of the transmission line phase.

\section{Joint-mode basis}
\label{app: joint modes basis}

A convenient basis to describe the physics of a coupled oscillator system is the joint-mode (or normal-mode) basis that diagonalizes the system Hamiltonian in the single-particle subspace. For a symmetric system composed of two identical oscillators, the joint modes'  position and momentum operators are defined as
\begin{equation}
\begin{aligned}
	 q_\pm & = (q_1\pm q_2)/\sqrt{2},\\
	 p_\pm & = (p_1\pm p_2)/\sqrt{2},
\end{aligned}
\end{equation}
corresponding to the common (+) and relative (-) oscillations of the two nanoparticles in the optical traps. In this new basis, Eq.~(1) from the main text decouples and we obtain two independent oscillators, described by the equations of motion
\begin{equation}
    \ddot{q}_\pm+\Omega_\pm^2q_\pm= \Omega_0 \sqrt{4\Gqba} n_\pm,
\end{equation}
which coincides with Eq.~(5) from the main text.
The effect of the coupling is twofold. On the one hand, we have new resonance frequencies (normal-mode splitting) defined as
\begin{equation}
\label{eq: normal mode splitting sup}
    \Omega_\pm^2=\Omega_0\left(\Omega_0 \pm\frac{4\Gamma\sqrt{\eta}\sin(\theta)}{1+\eta\mp2\sqrt{\eta}\cos(\theta)}\right).
\end{equation}
On the other hand, the optical force noise seen by each oscillator is modified by the coupling. This time, the two forces are uncorrelated $\langle \overline{n_+n_-}\rangle =0$ and their variance is
\begin{equation}
        \langle n_\pm^2\rangle = \mathcal{N}_{\pm}^2=\frac{1}{2}\frac{1-\eta}{1+\eta\mp2\sqrt{\eta}\cos(\theta)},
\end{equation}
as reported in Eq.~(7) of the main text.
The eigenfrequencies and the amplitude of the (non-local) fluctuations driving the two joint modes can be compactly written as 
\begin{equation}\label{eq: freqGqba_jointModes}
\begin{aligned}
	\Omega_{\pm}^2 & = \Omega_0^2 - 4\Gqba \Omega_0s_1,\\
	n_{\pm} & = g_\eta(s_1^{\pm} Y_{\pm} - s_2^{\pm} X_{\pm}),
\end{aligned}
\end{equation}
with coefficients
\begin{equation}\label{eq: s12_coeff}
	\begin{aligned}
		s_1^{\pm}&=\frac{\mp\sqrt{\eta}\sin\theta}{1+\eta \mp 2\sqrt{\eta}\cos\theta},\\
        s_2^{\pm}&=\frac{\eta \mp \sqrt{\eta}\cos\theta }{1+\eta \mp 2\sqrt{\eta}\cos\theta}.
	\end{aligned}
\end{equation}
In Eq.~\eqref{eq: freqGqba_jointModes}, we have defined the joint mode quadratures in analogy to the joint modes of the motional operators. It is worth also noticing that---in absence of damping---the dynamical stability of the oscillator dynamics requires that both joint mode frequencies are real-valued positive numbers. As $\Omega_+$ is real and positive in the entire parameter space $\theta \in \{0, \pi\}$, the system is dynamically stable if and only if $\Omega_{-}^2\geq 0$. The boundaries of the stability region are thus parametrized by the roots of
\begin{equation}
    \sin\theta+2\sqrt{\eta} A\cos(\theta)+ (1+\eta)A=0
\end{equation}
where $A=\Omega_0/(4\Gqba\sqrt{\eta})$. For angles $\theta\in [0,\pi]$, using some trigonometric relations and completing the squares, yields the solutions for the transmission line phase
\begin{equation}\label{eq_stability}
    \cos{\theta_{\pm}}=-\frac{2A^2\sqrt{\eta}(1+\eta)\pm\sqrt{1-A^2(\eta-1)^2}}{1+4A^2\eta}
\end{equation}
defining the unstable region through $\theta_{-}<\theta<\theta_{+}$. This instability can be understood as one of the modified spring constant of the normal modes becoming negative. The stability condition can be generalized in a rather straightforward way to the case including some mechanical damping $\gamma$ in the oscillator susceptivity, upon substituting $A$ with $A_{\gamma}=(\gamma^2/4+\Omega_0^2)/(4\Gqba\Omega_0\sqrt{\eta})$. 

\section{Transient solutions}

In the following, we calculate the evolution of the system vector state covariance matrix $\vb{\Sigma}(t) =: \langle \overline{\bm{v} \bm{v}^\mathrm{T}} \rangle$, where $\bm{v} = (q_+, p_+, q_-, p_-)^\mathrm{T}$ and where the overline denotes once again symmetrization. We start from an initial arbitrary state parametrized by $\vb{\Sigma}_0$. We assume a regime of motion governed by the photon recoil, such that we can safely neglect the effects of the thermal bath. We introduce the ratios between the bare oscillator frequency and the normal mode frequencies $r_\pm = \Omega_0/\Omega_\pm$, and write the quantum Langevin equations (QLEs) in the joint mode basis
\begin{subequations}
\label{eq: EOM joint basis}
\begin{align}
    \dot{q}_\pm &= \Omega_0 p_\pm, \\
    \dot{p}_\pm &= -\Omega_\pm r^{-1}_\pm q_\pm + \sqrt{4\Gqba} n_\pm.
\end{align}
\end{subequations}
Since the $\pm$ modes are decoupled and the fluctuating forces $n_{\pm}$ are uncorrelated, the evolution of the state vectors $\vb{v}_{\pm}=(q_{\pm},p_{\pm})^\mathrm{T}$ is independent of each other. We cast the QLEs in vector form $\dot{\vb{v}}_{\pm}=\vb{A}_{\pm}\vb{v}_{\pm}+\vb{w}_{\pm}$, introducing the drift matrix 
\begin{equation}
	\label{eq: evolution operator explicitely}
 \vb{A}_{\pm}=\begin{pmatrix}
		0 & \Omega_0\\
		-\Omega_{\pm}r^{-1}_{\pm} & 0 \\
	\end{pmatrix}
\end{equation} 
and the process noise vector $\vb{w}_{\pm}=\sqrt{4\Gqba}(0,n_{\pm})^\mathrm{T}$, fully characterized by the symmetrized covariance matrix
\begin{equation}
	\vb{W}_{\pm}=\langle \vb{w}_{\pm}\vb{w}_{\pm}^\mathrm{T} \rangle = 4\Gqba \mathcal{N}_{\pm}^2 \, \begin{pmatrix}
		0 & 0\\
		0 & 1\\
	\end{pmatrix}.
\end{equation}
Note that Eq.~\eqref{eq: evolution operator explicitely} can be decomposed into the form presented in the main text.
The evolution of the state vectors $\vb{v}_j(t)=\vb{\Phi}_j(t)\vb{v}_j(0) + \int_0^t\dd s \, \vb{\Phi}_j(t-s) \vb{w}_j(s)$ is ruled by the matrix exponential
\begin{align}
	&\vb{\Phi}_{j}(t) = e^{t \vb{A}_{j} } = \begin{pmatrix}
		\cos(\Omega_{j}t) & r_{\pm} \sin(\Omega_{j}t)\\
		-r_{\pm}^{-1} \sin(\Omega_{j}t) & 	\cos(\Omega_{j}t)
	\end{pmatrix} = \begin{pmatrix} \sqrt{r_\pm} & 0 \\ 0 & \frac{1}{\sqrt{r_\pm}}\end{pmatrix} \begin{pmatrix}
		\cos(\Omega_{j}t) & \sin(\Omega_{j}t)\\
		- \sin(\Omega_{j}t) & 	\cos(\Omega_{j}t)
	\end{pmatrix} \begin{pmatrix} \frac{1}{\sqrt{r_\pm}} & 0 \\ 0 & \sqrt{r_\pm}\end{pmatrix},
\end{align}
which we decomposed into the sequence of squeezing operations and rotation that we used in the main text. 
Similarly, the state vector covariance matrix evolves according to
\begin{equation}
	\begin{aligned}
		\vb{\Sigma}_j(t) & =  \vb{\Phi}_{j}(t)\vb{\Sigma}_0\vb{\Phi}_{j}(t)^\mathrm{T}+\int_0^t\dd s \,  \vb{\Phi}_{j}(t-s) \vb{W}_j \vb{\Phi}_{j}(t-s)^\mathrm{T}\\
		& = \vb{\Sigma}^{c}_j(t) + \vb{\Sigma}^{n}_j(t),\\
	\end{aligned}
\end{equation}
Where we separated the terms stemming from the unitary dynamics ($\vb{\Sigma}^{c}$) from those associated with decoherence ($\vb{\Sigma}^{n}$). Denoting the initial displacement variance, momentum variance, and their (symmetrized) covariance as $Q^2_0$, $P^2_0$, and $E_0$ respectively, the matrix elements of $\vb{\Sigma}^{c}$ read
\begin{equation}
\begin{aligned}
	\Sigma_{Q}^c & = Q^2_0\cos^2(\Omega_j t) + r_j^2 P^2_0\sin^2(\Omega_j t) + r_j E_0 \sin(2\Omega_j t)\\
	\Sigma_{P}^c & = P^2_0\cos^2(\Omega_j t) + r_j^{-2} Q^2_0\sin^2(\Omega_j t) - r_j^{-1} E_0 \sin(2\Omega_j t)\\
	\Sigma_{E}^c & = E_0 \cos(2\Omega_j t) + (r_j P_0^2-r^{-1}_j Q_0^2)\sin(2\Omega_j t)/2\\    
\end{aligned}
\end{equation}
while those of $\vb{\Sigma}^{n}$ are
\begin{equation}
\begin{aligned}
	\Sigma_{Q}^n & = 2\Gqba\Omega_j^{-1}\mathcal{N}^2_j r_j^2 [\Omega_j t - \sin(2\Omega_j t)/2]\\
	\Sigma_{P}^n & =   2\Gqba\Omega_j^{-1}\mathcal{N}^2_j [\Omega_j t - \Omega_j^{-1}\sin(2\Omega_j t)/2]\\
	\Sigma_{E}^n & =  2\Gqba\Omega_j^{-1}\mathcal{N}^2_j r_j \sin^2(\Omega_j t).\\    
\end{aligned}
\end{equation}
Notice that in the unstable regime these equations are valid, but as $\Omega_{-}$ becomes imaginary, the trigonometric functions turn into hyperbolic, yielding an exponential divergence of some of the expectation values of the state vector and of the covariance matrix elements.\\

Since the joint modes evolution is decoupled and uncorrelated, we can build the covariance matrix $\vb{\Sigma}(t)$ for the composite system state vector $\vb{v}^\mathrm{T} = (q_+\ p_+\ q_-\ p_-)$, using as diagonal blocks $\vb{\Sigma}_+(t)$ and $\vb{\Sigma}_-(t)$. The covariance matrix associated with the single particle basis $\vb{v}_{sp}^\mathrm{T} = (q_1\ p_1\ q_2\ p_2)$, can be obtained through $\vb{\Sigma}_{sp} = \bm{R} \bm{\Sigma} \bm{R}^\mathrm{T}$, with the transformation matrix
\begin{equation}
    \label{eq: transformation matrix}
    \bm{R} = \frac{1}{\sqrt{2}}\begin{pmatrix}
        1 & 0 & 1 & 0 \\
        0 & 1 & 0 & 1 \\
        1 & 0 & -1 & 0 \\
        0 & 1 & 0 & -1
    \end{pmatrix}.
\end{equation}
We can write write the covariance matrix in the form 
\begin{equation}
\vb{\Sigma}_{sp} = \begin{pmatrix} \bm{\alpha} & \bm{\gamma} \\ \bm{\gamma} & \bm{\beta}
    \end{pmatrix},
\end{equation}
where the blocks $\bm{\alpha}$ and $\bm{\beta}$ refer to the covariance matrices of the first and of the second particle respectively, while $\bm{\gamma}$ refers to cross correlations between the two particles. Due to the bilinear coupling among the nanoparticles motion, that amounts both to a beam-splitter and a two-mode squeezing interaction, we anticipate in the transiente dynamics of the system the build up of strong inter-particle correlations. We use as entanglement witness the minimum symplectic eigenvalue for our bipartite system
\begin{equation}
    \label{eq: entanglement criterion}
    \nu_\mathrm{min} = \sqrt{2\Delta - 2\sqrt{\Delta^2 - 4\mathrm{det}\bm{\Sigma}_{sp}}} < 1,
\end{equation}
with $\Delta = \mathrm{det}\bm{\alpha} + \mathrm{det}\bm{\beta} - 2\mathrm{det} \bm{\gamma}$.\\

In the main text, we use Eq.~\eqref{eq: entanglement criterion} to test whether an entangled state arises at some time $t$ from the coupled dynamics of the system. To initialize the system, we imagine to divert all the light present in the transmission lines onto a pair of homodyne receivers, thus suppressing the coupling, while recording measurements of the individual particle displacements. Such measurements can be used to stabilize with high fidelity the conditional state of the two particles. Specifically, we can extract the conditional state from the measurement record with the causal Wiener filter derived in Ref.~\cite{Meng2020}. Taking the limit of a negligible thermal decoherence, i.e. ($C_{q} = \Gqba/\gamma \gg 1$), this filter is characterized by a central frequency and linewidth given by
\begin{equation}\label{eq: WienerFilt_FreqLW}
    \begin{aligned}
        \Omega_W^2 & =\Omega_{0}^2\sqrt{1+\eta(4\Gqba/\Omega_0)^2},\\
        \Gamma_W & = \sqrt{2\Omega_W^2-2\Omega_{0}^2},
    \end{aligned}
\end{equation}
and produces a conditional state whose covariance matrix elements read
\begin{equation}\label{eq: CondState_CorrelationMatrix}
\begin{aligned}
    Q^{2}_W & =\frac{1}{8\eta}\frac{\Gamma_W}{\Gamma_{\pm}},\\
    P^{2}_W & =\frac{1}{8\eta}\frac{\Gamma_W}{\Gamma_{\pm}}\frac{\Omega_W^2}{\Omega_{\pm}^2},\\
    E_{W}   & =\frac{1}{16\eta}\frac{\Gamma_W^2}{\Gamma_{\pm}\Omega_{\pm}}.
\end{aligned}
\end{equation}
At $t=0$, up to moderate ratios $\Gqba/\Omega_0$, we assume that an optimal controller  can be used to prepare each particle into an unconditional state $\vb{\Sigma}_{1,2}\approx \vb{\Sigma}_{W}$. 

\section{In-loop position measurements of the joint modes }

\begin{figure}[t]
		\centering
		\includegraphics[trim=0cm 0cm 0cm 0cm, width=84mm]{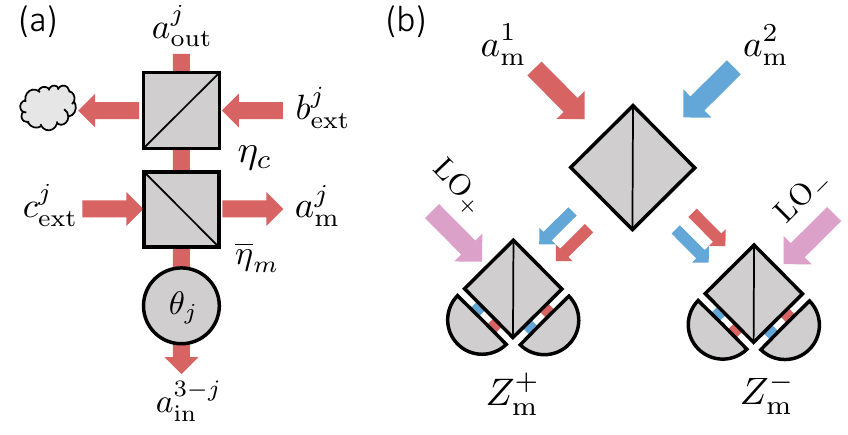}
		\caption{(a) Modeling of the modified transmission line allowing a fraction $\eta_\mathrm{m}$ of the collected light from the $j$-th nanoparticle to be routed onto a measurement apparatus. (b) Schematic illustration of the joint modes displacement measurement apparatus.}
		\label{fig:Meas_Scheme}
\end{figure}

In this section we describe a continuous measurement of the joint modes displacement amplitudes. To do so, we need to consider a slightly more complicated model for our transmission lines including three elements: two beamsplitters and a phase shifter, see Fig.~\ref{fig:Meas_Scheme}(a). The first beamsplitter models the information loss due to the finite collection efficiency and imperfect mode-matching into the transmission line, we denote its transmission $\eta_c$. The second beamsplitter is used to sample a small portion of the signal, and has transmission $\overline{\eta}_\mathrm{m}=1-\eta_\mathrm{m}$. The phase shifter allows to tune the transmission line phase delay $\theta$. The closure relations for the input fields in the optical loop are
\begin{equation}
\begin{aligned}
		a_\inn^{1,2}&=\expu^{\imu\theta}[\sqrt{\overline{\eta}_\mathrm{m}}(\sqrt{\eta_c}\aout^{2,1}+\sqrt{\overline{\eta_c}}b_\text{ext}^{2,1})+\sqrt{\eta_\mathrm{m}}c_\text{ext}^{2,1}]\\
		&=\expu^{\imu\theta}[\sqrt{\eta}\aout^{2,1}+\sqrt{\overline{\eta}}a_\text{ext}^{2,1}],
\end{aligned}
\end{equation}
where we used the fact that $b^{j}_\text{ext}$ and $c^{j}_\text{ext}$ are uncorrelated noise sources and defined the line transmission $\eta=\eta_c\overline{\eta}_\mathrm{m}$. The above equation tells us that the input fields in Eq.~(4) are not modified. Nevertheless we need to keep in mind the noise decomposition 
\begin{equation}\label{eq:aext_noiseDecomposition}
	a_\text{ext}^{j}=\sqrt{\frac{\overline{\eta}_\mathrm{m}\overline{\eta}_c}{\overline{\eta}}}b_\text{ext}^{j}+\sqrt{\frac{\eta_\mathrm{m}}{\overline{\eta}}}c_\text{ext}^{j}
\end{equation}
in order to take into account correlations among noise sources when closing a feedback loop.\\

The optical fields on the reflection ports of the pick-off beamsplitters read
\begin{equation}
	a_\text{m}^{j}= -\sqrt{\eta_c\eta_\mathrm{m}}\aout^{j}-\sqrt{\overline{\eta}_c\eta_\mathrm{m}}b^{j}_\text{ext}+\sqrt{\overline{\eta}_\mathrm{m}}c^{j}_\text{ext}.
\end{equation}
Inserting the input-output relations [Eqs.~(2) and~(4)] from the main text one finds the coherent part of the fields 
\begin{equation}
	a^{1,2}_\text{m,c}=-\imu \frac{\sqrt{2\Gamma_\text{m}}}{1-\alpha^2}[q_{1,2}+\alpha q_{2,1}],
\end{equation}
with $\Gamma_\text{m}=\eta_c\eta_\mathrm{m}\Gqba$ and with the noise terms
\begin{equation}
\begin{aligned}
	a^{1,2}_\text{m,n} & = -\sqrt{\eta_\mathrm{m}\overline{\eta}_c}\left[\frac{1}{1-\alpha^2} \right]b_\text{ext}^{1,2} +\sqrt{\overline{\eta}_\mathrm{m}}\left[1-\frac{\eta_\mathrm{m}}{\overline{\eta}_\mathrm{m}}\frac{\alpha^2}{1-\alpha^2} \right]c_\text{ext}^{1,2} -\frac{\alpha \sqrt{\eta_\mathrm{m}}}{1-\alpha^2}\left[ \sqrt{\overline{\eta}_c}b_\text{ext}^{2,1} + \sqrt{\frac{\eta_\mathrm{m}}{\overline{\eta}_\mathrm{m}}}c_\text{ext}^{2,1} \right].
\end{aligned}
\end{equation}
The fields $a^{j}_\text{m}=a^{j}_\text{m,c}+a^{j}_\text{m,n}$ carry information on particle displacements. Depending on the transmission line phase $\theta=\text{arg}(\alpha)$ the displacements may be encoded in different quadratures, yielding a sub-optimal measurement of $q_{1,2}$. If we instead mix these two fields on a beamsplitter, as shown in Fig.~\ref{fig:Meas_Scheme}(b), the coherent part of the fields at the symmetric (+) and anti-symmetric (-) output ports provide a direct measurement of the joint modes
\begin{equation}
	a^{\pm}_\text{m,c}=-\imu \frac{\sqrt{2\Gamma_\text{m}}}{1\mp \alpha}q_\pm,
\end{equation}
while the noise terms read
\begin{equation}
	a^{\pm}_\text{m,n}  =-\frac{\sqrt{\eta_\mathrm{m}\overline{\eta}_c}}{1\mp \alpha}b_\text{ext}^{\pm} + \sqrt{\overline{\eta}_\mathrm{m}}\left[1\mp \frac{\eta_\mathrm{m}}{\overline{\eta}_\mathrm{m}}\frac{\alpha}{1\mp \alpha} \right] c_\text{ext}^{\pm},
\end{equation}
with $b_\text{ext}^{\pm}=(b_\text{ext}^{1}\pm b_\text{ext}^{2})/\sqrt{2}$ and $c_\text{ext}^{\pm}=(c_\text{ext}^{1}\pm c_\text{ext}^{2})/\sqrt{2}$.\\

A phase quadrature measurement of $a_\text{m}^{\pm}$ yields 
\begin{equation}\label{eq: Ym_value}
	\begin{aligned}
		Y_\text{m}^{\pm}& = \sqrt{4\Gamma_\text{m}} s^{\pm}_c q_{\pm} + \sqrt{\eta_\mathrm{m}\overline{\eta}_c} [ s_1^{\pm} X_b^{\pm} + s_c^{\pm} Y_b^{\pm}] + \sqrt{\overline{\eta}_\mathrm{m}}\frac{\eta_\mathrm{m}}{\overline{\eta}_\mathrm{m}}\left[ s_1^{\pm} X_c^{\pm} + \left( \frac{\overline{\eta}_\mathrm{m}}{\eta_\mathrm{m}} + s_2^{\pm}\right) Y_c^{\pm} \right],
	\end{aligned}
\end{equation}
while an amplitude measurements yields
\begin{equation}\label{eq: Xm_value}
	\begin{aligned}
		X_\text{m}^{\pm}& = -\sqrt{4\Gamma_\text{m}} s^{\pm}_1 q_ {\pm} + \sqrt{\eta_\mathrm{m}\overline{\eta}_c} [ s_c^{\pm} X_b^{\pm} - s_1^{\pm} Y_b^{\pm}] + \sqrt{\overline{\eta}_\mathrm{m}}\frac{\eta_\mathrm{m}}{\overline{\eta}_\mathrm{m}}\left[ \left( \frac{\overline{\eta}_\mathrm{m}}{\eta_\mathrm{m}} + s_2^{\pm}\right) X_c^{\pm} - s_1^{\pm} Y_c^{\pm} \right],
	\end{aligned}
\end{equation}
where we have introduced the coefficient
\begin{equation}
		s_c^{\pm}=\frac{\pm\sqrt{\eta}\cos\theta - 1}{1+\eta \mp 2\sqrt{\eta}\cos\theta},
\end{equation}
while $s_{1,2}$ are defined by Eq.~\eqref{eq: s12_coeff}. For a fixed analyzer angle $\varphi$ one measures the quantity $Z_\text{m}^{\pm}=\cos\varphi X_\text{m}^{\pm}+\sin\varphi Y_\text{m}^{\pm}$. In general, the measurement outcomes can be decomposed into signal and imprecision terms $Z_\text{m}^{\pm}=Z_\text{sig}^{\pm} + Z_\text{imp}^{\pm}$. The amount of information gathered on the joint modes position for a given analyzer angle $\varphi$ reads 
\begin{equation}\label{eq: MeasurementStrength}
    Z_\text{sig}^{\pm}=\sqrt{4\Gamma_\text{m}} \left(\sin\varphi s_c^{\pm}-\cos\varphi s_{1}^{\pm}\right) q_{\pm}.
\end{equation}
In the limiting case where all the light is diverted onto the detectors ($\eta_\mathrm{m}=1$), the particles decouple, and all the information on the particles position is encoded in the phase-quadrature of the measurement field. Furthermore, only the vacuum fluctuations amplitude quadratures $X_{b,c}^{\pm}$ drive the particle motion, while the phase quadratures only enter the imprecision noise terms in $Z_\text{m}^{\pm}$. This is not surprising, as such a limiting case, corresponds to the standard framework of continuous position measurements in free-space optomechanics.\\

However, beyond this limiting case, the amplitude and phase quadratures of the vacuum fields contribute to both the imprecision $Z_\text{m}^{\pm}$ and measurement back-action $n_{\pm}$ terms, indicating a correlation between these quantities. The question then, is whether for some value of the analyzer angle we can find a measured field quadrature that provides the maximal information on the joint mode displacement without being affected by bath correlations. We calculate the (symmetrized) correlator
\begin{equation}\label{eq: ZN_correlation}
\begin{aligned}
	\mathcal{C}_{\varphi} &=\langle \overline{ n_{\pm} Z_\text{imp}^{\pm}} \rangle \\
	& = g_{\eta}  \langle \overline{(s_1^{\pm} Y_{\pm} - s_2^{\pm} X_{\pm})(\cos\varphi X_\text{m,n}^{\pm}+\sin\varphi Y_\text{m,n}^{\pm})} \rangle\\
	&=\sqrt{\eta_c\eta_\mathrm{m}}(\sin\varphi s^{\pm}_c + \cos\varphi s^{\pm}_1),
\end{aligned}	
\end{equation}
where we plugged the external field decomposition \eqref{eq:aext_noiseDecomposition} in $n_{\pm}$ and denoted $X_\text{m,n}^{\pm}$ and $Y_\text{m,n}^{\pm}$ the fluctuating terms in Eq.~\eqref{eq: Xm_value} and \eqref{eq: Ym_value}, respectively. \\

Remarkably, from the $\varphi$ dependence of Eq.~\eqref{eq: MeasurementStrength} and \eqref{eq: ZN_correlation} becomes clear that whenever Eq.~\eqref{eq: ZN_correlation} vanishes for some $\tilde{\varphi}$,  Eq.~\eqref{eq: MeasurementStrength} is at an extremal (maximal) value. The optimum angle for which one gathers the maximal information on the joint mode displacement is parameterized by the expression
\begin{equation}\label{eq: OptimumAnalizerAngle}
    \tan\tilde{\varphi}_\pm=-s_c^{\pm}/s_1^{\pm}.
\end{equation}
Inserting the above expression in Eq.~\eqref{eq: MeasurementStrength}, one obtains the optimum measurement amplitude
\begin{equation}\label{eq: OptimalMeasurementSignal}
\begin{aligned}
    Z_\text{opt} & = \sqrt{4\Gamma_\text{m}} [\sin\tilde{\varphi}\, s_c^{\pm}-\cos\tilde{\varphi}\, s_{1}^{\pm}] q_{\pm}\\
    &=\mp \sqrt{4\Gamma_\text{m}[(s_1^{\pm})^2+(s_c^{\pm})^2]}q_{\pm}\\
    &=\mp \sqrt{\frac{4\Gamma_\text{m} }{1+\eta\mp 2\sqrt{\eta}\cos(\theta)}}\, q_{\pm}.
\end{aligned}
\end{equation}
In absence of correlations we can also define the effective measurement efficiency at the optimal measurement quadrature
\begin{equation}\label{eq: measEff_OptimalMeasQuadrature}
    \tilde{\eta}_{\pm}=\frac{\Gamma_\text{m}(s_c^2+s_1^2)q^2_\text{zpf}}{\Gamma_\text{n} \langle n_{\pm}^2 \rangle}=\frac{\eta\eta_\mathrm{m}}{\overline{\eta}\overline{\eta}_\mathrm{m}}.
\end{equation}
Interestingly, the above expression depends only on the measured fraction $\eta_\mathrm{m}$ and on the transmission line collection efficiency $\eta$ and in the limit $\eta_c\to 1$ yields  $\tilde{\eta}_{\pm}\to1$ regardless of the choice of $\eta_\mathrm{m}$.\\ 

As the joint modes dynamics is uncorrelated, and at the optimal analyzer angle the measured field properties can be mapped onto the standard framework of continuous position measurements, we can resort to the standard optomechanics toolbox. We can use again the causal Wiener filters defined in App.~C to write down the joint modes conditional state covariance matrix. This can be done upon operating the substitutions $\Omega\to\Omega_{\pm}$, $\Gqba\to 2\Gqba\mathcal{N}^2_{\pm}$ and $\eta\to \tilde{\eta}_{\pm}$ in Eq.~\eqref{eq: WienerFilt_FreqLW} and \eqref{eq: CondState_CorrelationMatrix}. \\

In the main text we use the resulting expressions to obtain an analytic expression for the entanglement witness \eqref{eq: entanglement criterion}, see Fig.~5 in main text. We conclude that the conditional state accessible by continuously monitoring the optimal optical field quadrature (cf. Eq.~\eqref{eq: OptimalMeasurementSignal}) can be used to stabilize an entangled steady-state of motion of the two nanoparticles.

\section{Entanglement verification}

So far, we have seen that we are able to generate an entangled state by coupling the nanoparticles' motion via an optical loop. In this appendix, we try to answer the following question: how do we certify this state?

Let's assume we have generated an entangled state by the proposed protocol. The state is characterized by a covariance matrix $\vb{\Sigma}$ and the entanglement witness is $\nu_\text{min}[\vb{\Sigma}]$.
At a given time $t=0$, which we refer to as the initial time later on, we decoupled the two nanoparticles by opening the loop.
In addition, we start measuring the output fields according to the scheme shown in Fig.~\ref{fig:Meas_Scheme}(b) (with $\eta_\mathrm{m}=1$).
Both nanoparticles are now subjected to a continuous position measurement, the quantum backaction of which destroys the initial state.
Nevertheless, if we record the measurements outcomes, we can extract from them by appropriate filtering (i.e., retrodiction) the best estimates of the nanoparticles' position and momentum \textit{at the initial time}, which we can then use to calculate the entanglement witness.
These estimates are subjected to both systematic and statistical errors, respectively due to the nature of the measurement scheme and to the finite size of the ensemble of experimental realizations.
In this appendix, we will first show how the dynamics evolves under a continuous position measurement. Then, we will summarize the theory of retrodiction and how to estimate the system's degrees of freedom from the outcomes. Finally, we assess the errors in the entanglement witness for a typical experimental scenario.

From the initial time on, the nanoparticles dynamics in the Heisenberg picture is described by the following quantum Langevin equations
\begin{equation}\label{eq: QLE_singleparticle}
    \begin{aligned}
        \dot{q}_j &=\Omega_0 p_j,\\
        \dot{p}_j &=-\Omega_0 q_j +\sqrt{4\Gqba}X^{j}_\text{in},
    \end{aligned}
\end{equation}
where $j=+,-$. 
We notice that there is no coupling between the two nanoparticles at this point. Nevertheless we keep working on the joint mode basis only for simplicity.
These equations can be rewritten more compactly as $\dot{\vb{x}}=\vb{A} \vb{x} + \vb{w}$, where $\vb{x}=\left(q_+,p_+,q_-,p_-\right)^\mathrm{T}$  is a state vector, $\vb{w}=\left(0,\sqrt{4\Gqba}X^{+}_\inn,0,\sqrt{4\Gqba}X^{-}_\inn\right)^\mathrm{T}$ is a  force noise vector with autocorrelation matrix
\begin{equation}
    \vb{W}=\langle\overline{\vb{w}(t)\vb{w}(t')^\mathrm{T}} \rangle=\begin{pmatrix}
        0 & 0 & 0 & 0\\
        0 & 2\Gqba & 0 & 0\\
        0 &0 &0&0\\
        0&0&0&2\Gqba
    \end{pmatrix}\delta(t-t')
\end{equation}
and the drift matrix
\begin{equation}
    \vb{A}=\begin{pmatrix}
        0 & \Omega_0&0&0\\
        -\Omega_0 & 0 & 0 & 0\\
        0 &0 &0 &\Omega_0\\
        0&0&-\Omega_0&0
    \end{pmatrix}.
\end{equation}

The measurement outcomes are represented by the photocurrent vector $\vb{z}=\left(z_{+},z_{-}\right)^\mathrm{T}$, which is connected to the state vector according to $\vb{z} = \vb{C}^\mathrm{T} \vb{x} + \vb{v} $, where
\begin{equation}
    \vb{C}^\mathrm{T}=\sqrt{4\eta_c\Gqba}\begin{pmatrix}
        1 & 0 & 0 & 0\\
        0 & 0 & 1 & 0
    \end{pmatrix}
\end{equation}
describes the measured observables, $\eta_c$ is the detection efficiency and $\vb{v}=\left(Y^{+}_\text{imp},Y^{-}_\text{imp}\right)^\mathrm{T}$, with $Y^{\pm}_\text{imp}=\sqrt{\eta_c}\,Y^{\pm}_\text{in}+ \sqrt{\overline{\eta}_c}\,Y^{\pm}_\text{n}$, is the measurement noise vector with autocorrelation matrix
\begin{equation}\label{eq:measurement_noise_matrix}
    \vb{V}=\langle\overline{\vb{v}(t)\vb{v}(t')^\mathrm{T}} \rangle=\begin{pmatrix}
        1/2 & 0\\
        0 & 1/2
    \end{pmatrix} \delta(t-t').
\end{equation}
We notice that the force noise and the measurement noise are uncorrelated when we measure the optical phase quadrature.

In the long-time limit, the state approaches a thermal state, with no information about the initial state, which is decayed away.
To extract some information about the initial state $\vb{x}_0$, we need to retrodict the state vector based on future outcomes.
We refer to the retrodicted state vector as $\vb{x}_E$. The uncertainty of the retrodicted state vector is encoded in the conditional covariance matrix $\vb{V_E}$.

The theory of retrodicted measurements allows us to extract these two quantities from the measured outcomes and the known system parameters.
In fact, we can write the retrodicted state vector in terms of the outcomes according to
\begin{equation}\label{eq:retrodicted_mean_values}
-\dd \vb{x_E} = \vb{M_E} \vb{x_E} \dd t + \left(2 \vb{V_E} \mathrm{Re}(\vb{C}) + \sigma \mathrm{Im}(\vb{C})\right)\vb{z}\dd t,
\end{equation}
where $-\dd\vb{x_E}=\vb{x_E}(t-\dd t)-\vb{x_E}(t)$, the modified drift matrix is 
\begin{equation}
\vb{M_E} = -\vb{A}-2\sigma\mathrm{Im}(\vb{C})\mathrm{Re}(\vb{C}^\mathrm{T})-4\vb{V_E}\mathrm{Re}(\vb{C})\mathrm{Re}(\vb{C}^\mathrm{T})
\end{equation}
and $\sigma$ is 
\begin{equation}
    \sigma=\begin{pmatrix}
        0& 1 &0&0\\
        -1 &0 & 0 & 0\\
        0 & 0&0&1\\
        0&0&-1&0
    \end{pmatrix}.
\end{equation}
The conditional covariance matrix, instead, follows the deterministic Riccati equation
\begin{equation}\label{eq:riccati_backward}
-{\vb{\dot{V}_E}} = \vb{M_E}\vb{V_E} + \vb{V_E}\vb{M_E}^\mathrm{T} + \vb{W} + 4\vb{V_E}\mathrm{Re}(\vb{C})\mathrm{Re}(\vb{C}^\mathrm{T})\vb{V_E},
\end{equation}
where $-{\vb{\dot{V}_E}}=\left(\vb{V_E}(t-\dd t)-\vb{V_E}(t)\right)/\dd t$.
For a retrodiction time longer than $1/(\eta_c\Gqba)$, the covariance matrix $\vb{V_E}$ approaches its steady-state value $\vb{V}^\infty_E$.
Equation~\eqref{eq:retrodicted_mean_values} can be approximated by a finite difference equation, which can be used to compute numerically the retrodicted state vector $\tilde{\vb{x}}_E$ for a given measurement outcome $\tilde{\vb{z}}$. 
Here we indicate with $\tilde{\cdot}$ quantities extracted from finte data rather than theoretical ones.

\begin{figure}[t]
		\centering
		\includegraphics[trim=0cm 0cm 0cm 0cm, width=100mm]{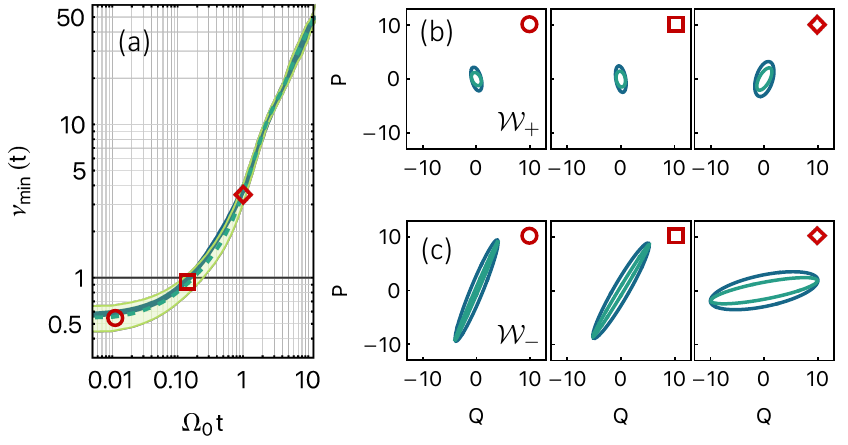}
		\caption{(a) The solid line traces the time evolution of the state vector negativity after the optical loop has been severed at $t=0$, for $\Gamma=2\Omega_0$ and $\eta=1/2$. The dashed line indicates the retrodicted negativity for the estimator state. The negativity estimate $2\sigma$ confidence level is shaded in gray considering averaging over a thousand trajectories. (b,c) Snapshots of the symmetric (top panels) and anti-symmetric (bottom panels) joint mode state (blue) and estimate (aquamarine) vector covariance matrix $1\sigma$ ellipses at $\Omega_0 t=(0.01,0.15,1.0)$. The fact that the estimated confidence ellipses always enclose those of the state vector is a signature of the inevitable added noise due to the measurement process.  }
		\label{fig:Entanglement_verification}
\end{figure}
We can propagate the finite difference equation from the last acquired point to the initial time $t=0$ to estimate the initial state vector, $\tilde{\vb{x}}_{E,0}$.
Then, we use this ensemble to calculate the covariance matrix $\tilde{\vb{\Sigma}}_{E, 0} = \langle \overline{\tilde{\vb{x}}_{E,0} \, \tilde{\vb{x}}_{E,0}^\mathrm{T}}\rangle$.
In practice, the experimenters have at their disposal a finite set of realizations of the measurement outcomes $\tilde{\vb{z}}$.
This introduces a \textit{statistical error} in  $\tilde{\Sigma}_0$ which depends on the ensemble size.
In addition, the retrodicted state vector $\tilde{\vb{x}}_{E,0}$ contains an additional error which depends on the measurement scheme. For instance, an ideal weak continuous position measurement adds half zero-point unit in both position and momentum.
We refer to this uncertainty as \textit{systematic error}.

These errors in the covariance matrix propagates in the final entanglement witness. 
We now investigate quantitatively their impact.

The systematic error arises from the conditional covariance matrix according to $\tilde{\vb{\Sigma}}_E=\tilde{\vb{\Sigma}} + \vb{V}_E$.
This conditional covariance matrix depends on the system parameters $(\Omega_0, \Gqba, \eta_c)$ and can be subtracted from the estimated $\tilde{\vb{\Sigma}}_E$ if they are known exactly.
In practice, the experimenters know them only within a finite precision.
Let's denote by $(\sigma_\Omega, \sigma_{\Gqba}, \sigma_\eta)=(0.1\%, 5\%, 5\%)$ the relative errors of these parameters and assume them uncorrelated from each other. The values we have assumed are for typical experimental scenario.
By propagating these errors, we find that the entanglement witness $\tilde{\nu}_\text{min}$ calculated on the matrix $\vb{V}_E$ has a $2\sigma$-confidence level of $7\%$.

Let's focus now on the statistical error.
To estimate it, we simulate $10^3$ measurement outcomes $\tilde{\vb{z}}$ according to Eqs.~\eqref{eq: QLE_singleparticle}-\eqref{eq:measurement_noise_matrix}.
We use a stochastic Runge-Kutta algorithm of order $3/2$ with a time-step $\Omega_0 \Delta t = 10^{-2}$.
The initial conditions are drawn from a probability distribution function which equals the Wigner function corresponding to the initially entangled state.
We apply the discrete version of Eq.~\eqref{eq:retrodicted_mean_values} to the simulated outcomes in order to obtain an ensemble of retrodicted state vectors $\tilde{\vb{x}}_{E}$, which we use to construct the matrix $\tilde{\vb{\Sigma}}_E$.
We calculate the uncertainties in the elements of this matrix by means of a bootstrapping technique.
Then, we propagate these uncertainties to the entanglement witness calculated $\tilde{\nu}_\text{min}$ calculated on this matrix.
Finally, we add in quadrature the statistical uncertainty of $\tilde{\nu}_\text{min}[\tilde{\vb{\Sigma}}_E]$ and the systematic one of $\tilde{\nu}_\text{min}[\vb{V}_E]$.

In Fig.~\ref{fig:Entanglement_verification}(a), we compare the theoretical entanglement witness $\nu_\text{min}[\vb{\Sigma}]$ with the one from the retrodicted measurements, $\tilde{\nu}_\text{min}$, at different times. The shaded area marks the $2\sigma$-confidence interval due to both systematic and statistical errors, as discussed above. 
We notice that the confidence interval is below the threshold of $1$, which allows us to successfully certify entanglement.
In panels (b), (c) we also show covariance ellipse of the state and of its reconstruction.